\documentclass[aps,twocolumn,showpacs,amsmath,amssymb,pre]{revtex4}
\usepackage{epsfig,graphicx}

\begin{document}
\title{Multi-resonance of energy transport and absence of heat pump in a force-driven lattice}

\author{Song Zhang$^{1}$}
\author{Jie Ren$^{2,1}$}
\email{renjie@nus.edu.sg}
\author{Baowen Li$^{1,2}$}
\email{phylibw@nus.edu.sg}

\affiliation{$^1$Department of Physics and Centre for Computational Science and Engineering,
National University of Singapore, Singapore 117546, Republic of Singapore \\
$^2$NUS Graduate School for Integrative Sciences and Engineering,
Singapore 117456, Republic of Singapore}

\begin{abstract}
Energy transport control in low dimensional nano-scale systems has
attracted much attention in recent years. In this paper, we
investigate the energy transport properties of Frenkel-Kontorova
lattice subject to a periodic driving force, in particular,  the resonance behavior of the energy current by varying the
external driving frequency. It is discovered that in certain parameter ranges,
multiple resonance peaks, instead of a single resonance, emerge. By
comparing the nonlinear lattice model with a harmonic chain, we
unravel the underlying physical mechanism for such resonance
phenomenon. Other parameter dependencies of the resonance behavior
are examined as well. Finally, we demonstrate that heat pumping is
actually absent in this force-driven model.
\end{abstract}
\pacs{05.60.-k, 44.10.+i, 66.70.-f, 07.20.Pe}


\date{\today}
\maketitle

\section{Introduction}

In the last decade much attention has been paid to heat conduction
on the nanoscale. Numerous studies in the quest of manipulating heat
have brought substantial progresses in the area of \emph{Phononics},
the science and engineering of phonons~\cite{WL08}. Theoretical
models of various thermal devices such as thermal
rectifier~\cite{devices1}, thermal transistor~\cite{devices2},
thermal logic gates~\cite{devices3} and thermal
memory~\cite{devices4} have been proposed to control phonon-based
thermal transport. In addition, some experimental works have been
carried out. For instance, solid-state thermal diodes based on
asymmetric nanotube \cite{exp_diode1}, asymmetric cobalt
oxides~\cite{exp_diode3}, the nanotube phonon
waveguide~\cite{waveguide} and solid thermal memory\cite{Xie2011} have been realized experimentally. In the
works mentioned above, heat always flows from high temperature
region to low temperature one. This is in accordance with the
second law of thermodynamics.

However, just like its macroscopic counterpart, a heat
pump which directs heat against thermal bias by using external force exists on the molecular
level. Recent studies have suggested several such kind of models based on different
mechanisms. Li \emph{et al}.
proposed a heat ratchet to direct heat flux from one bath to another
in a nonlinear lattice, which periodically adjusts two baths'
temperatures while the average remains equal~\cite{Li0809}. Later,
Ren and Li demonstrated that heat energy can be rectified between
two baths of equal temperatures at any instant and the correlation
between the baths can even direct energy current against thermal
bias without external modulations~\cite{Ren10}. Beyond those
classical models, a quantum heat pump consisting of a molecule
connected to two phonon baths was proposed
recently~\cite{Segal0608}, and the Berry phase effect induced heat
pump was unraveled as well~\cite{Ren10PRL}.

Nevertheless, some contradiction exists in the force-driven heat
pump. Marathe \emph{et al} showed that under periodic force driving,
two coupled oscillators connected with thermal baths fail to function
as a heat pump~\cite{Marathe07}. Later Ai \emph{et al} claimed the
heat pumping appeared in Frenkel-Kontorova (FK) chain under
the influence of a periodic driving force \cite{Ai2010}. In addition,
they observed a thermal resonance phenomenon that the heat flux
attained a maximum value at a particular driving frequency, which is
similar to the previously found resonance induced by driving bath
temperatures~\cite{Ren10}. However, other than the temperature
driven case, a clear physical mechanism of the driving-force-induced
resonance is still unavailable.

In this paper, we investigate the force-driven FK chain as a typical
nonlinear lattice model and analyze thermal properties which are
frequency dependent. We discover that multiple
resonance peaks, instead of a single peak, emerge in certain
parameter ranges. The origin of this phenomenon in fact relies on
the eigenfrequencies of this system. The paper is organized as
follows.  First of all, we will introduce the FK model and briefly
show the crossover from a single resonance to multiple resonances.
To look into the underlying mechanism of the multi-resonance
phenomenon, we invoke a harmonic model through which analytic
expressions of various quantities are obtained. Then the resonance
behavior of the FK model is discussed in detail. We will show that as
far as the resonance property is concerned, there is much similarity
between the FK model and the harmonic model. The effect of different
system parameters on the resonance phenomenon is explored as well.
In the end, we shall briefly demonstrate that the force-driven model
fails to perform as a heat pump.

\section{Model and Results}
\begin{figure}\vspace{-10mm}
\scalebox{0.4}[0.3]{\hspace{10mm}\includegraphics{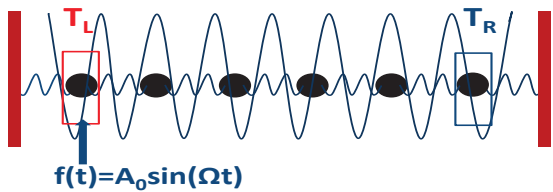}}\vspace{-20mm} 
\caption{ \label{fig1FK} (Color online) Schematic of a
one-dimensional FK lattice coupled to two Langevin baths. The left end is
attached to a periodic
driving force $f(t)$.} 
\end{figure}
\begin{figure}
\includegraphics[scale=0.36]{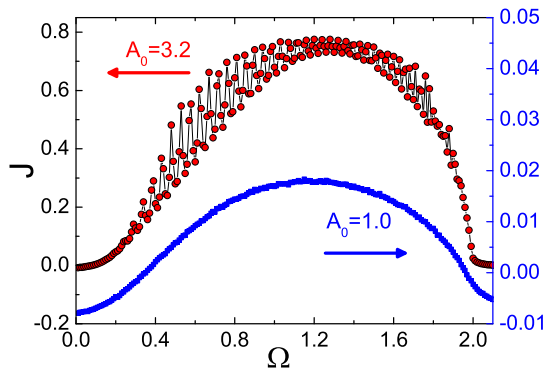}
\caption{(Color online) Energy current $J$ vs the driving frequency
$\Omega$ in the force-driven FK model. Parameters are $T_L=0.4, T_R=0.6,
k=1, V=5, \gamma=1$, and $N=64$. $J$ denotes $J_i$ ($i>\alpha=1$).
The left arrow means the upper curve corresponds to the left scale
while the right arrow means the lower curve corresponds to the right
scale. At large $\Omega$, the response of the system cannot catch up
with the fast driving as if there is no driving. At small $\Omega$,
the system reduces to a quasi-static counterpart as if there is no
driving as well. Therefore, at these two regions, $J<0$ since
$T_L<T_R$.}\label{fig:MRP}
\end{figure}

We start with a FK chain coupled to Langevin heat baths at two ends.
A periodic force is applied to the left end (See Fig.~\ref{fig1FK}).
The Hamiltonian of the system is given as:
\begin{align}\label{Hamil}
H&=\sum^N_i
\frac{p_i^2}{2m}+\frac{1}{2}k(x_i-x_{i+1})^2-\frac{V}{(2\pi)^2}\cos(\frac{2\pi}{a}
x_i)\qquad\nonumber \\
&\ \quad -\delta_{i\alpha}x_{i}f(t), 
\end{align}
where $f(t)=A_0\sin(\Omega t)$ is the driving force with amplitude
$A_0$ and frequency $\Omega$ imposed on particle $\alpha$. $x_i$
denotes the displacement of the $i$th particle and $p_i$ the
corresponding momentum. $k$ is the spring constant, $V$ the strength
of the on-site potential and $a=1$ the spacing between adjacent
particles. We set $\alpha=1$ in follows unless otherwise stated.
Under fixed boundary condition, the equation of motion (EOM) of the
coupled particles ($i=2,\cdots N-1$) reads:
\begin{subequations}
    \begin{align}
m\ddot x_1&=-\frac{V}{2\pi}\sin(2\pi
    x_1)+k(x_2-2x_1)+f(t)\qquad\\
    &\qquad-\gamma\dot x_1\ +\eta_L(t), \nonumber\\
m\ddot x_i&=-\frac{V}{2\pi}\sin(2\pi
    x_i)+k(x_{i+1}+x_{i-1}-2x_1), \quad \\
m\ddot x_N&=-\frac{V}{2\pi}\sin(2\pi
    x_N)+k(x_{N-1}-2x_N)\qquad\nonumber\\
    &\qquad-\gamma\dot x_N\ +\eta_R(t).
    \end{align}
\end{subequations}
The two noise terms are white noise with
$\langle\eta_i(t)\eta_j(t')\rangle=2\gamma k_BT_i\delta_{ij}\delta(t-t')$,
$(i,j=L,R)$, where $\gamma$ is the friction coefficient, $k_B$ is
the Boltzmann's constant and $T_{L(R)}$ denotes the temperature of
$L(R)$ reservoir. Without loss of generality, we set $m=1$, $k_B=1$
and integrate the EOM by the symplectic Velocity Verlet algorithm.
The simulation are performed long enough (of order $10^8$) with a
time step of $0.005$ to allow the system to reach a non-equilibrium
steady state. The local energy current is defined as $I_i(t)=
\langle-k\dot x_i(x_i-x_{i-1})\rangle$, where $\langle\cdots\rangle$
denotes the ensemble average. 
After the transient time, the transport approaches an oscillatory
steady state and the periodic average of local current is equivalent at each
site $i$, reading
\begin{equation}
J_i=\frac{\Omega}{2\pi}\int_0^{{2\pi}/{\Omega}}I_i(t)dt.
\end{equation}
There is no doubt that the thermal resonance exists for the energy
current as a function of the driving frequency. However,
surprisingly, we observe that multiple resonance peaks emerge when
the amplitude $A_0$ is increased, as shown in Fig.~\ref{fig:MRP}. To
analyze this multi-resonance phenomenon further, we invoke the
harmonic chain model, of which clear analytic results are possible.

\subsection{Analytic results for force-driven harmonic lattice}\label{sec:A}
Harmonic potential is a second order approximation of realistic
potential around its minimum. Under harmonic approximation,
phonons will not interact with each other. Starting from this
noninteracting picture, interactions within phonon gas can be
introduced, by including higher order terms in the potential
expansion. Compared to non-linear models such as the FK lattice, a
harmonic model has the advantage of having an exact solution and it
gives many satisfactory explanations about thermal properties of the
crystal.

The linear harmonic model considered in this section is obtained by
replacing the nonlinear on-site potential in Eq.~\eqref{Hamil} by
$\frac{1}{2}k_ox_i^2$, where $k_o$ is the force constant of the
on-site harmonic potential. As we shall see, under certain
conditions the effect of the driving force can be separated from
that of the thermal reservoirs. The EOM in a compact matrix form is:
\begin{equation}\label{EOM_dr}
\mathbf{M}\ddot X=-\mathbf{\Phi} X -\mathbf{\Gamma} \dot X +
\eta(t)+F(t),
\end{equation}
where $\mathbf{M}$ is the mass matrix set as identity matrix and
$\mathbf{\Phi}$ the real symmetric force matrix, with
$\Phi_{ij}=(2k+k_o)\delta_{ij}-k\delta_{i,j+1}-k\delta_{i,j-1}$.
$\Gamma_{ij}=\gamma\delta_{ij}(\delta_{i1}+\delta_{iN})$ denotes the
coupling to reservoirs. $X=(x_1, x_2, \cdots, x_N)^T$ is the vector
of displacements and $\eta(t)=(\eta_L, 0, \cdots, 0, \eta_R)^T$
depicts the thermal noise in reservoirs. $F(t)$ is a column vector
with elements $f_i(t)=\delta_{i\alpha}A_0\sin(\Omega t)$, denoting
the driving force which acts on the $\alpha$th particle. Defining
the Fourier transform of quantity $A$ and its inverse:
$$\tilde{A}(\omega)=\frac{1}{2\pi}\int_{-\infty}^{\infty} e^{i\omega
t}A(t) dt,\quad{} A(t)=\int_{-\infty}^{\infty} e^{-i\omega t}\tilde
A(\omega)dw,$$ and applying them to Eq.~\eqref{EOM_dr}, we obtain
\begin{eqnarray}\label{X_D}
X(t)&=&X^s(t)+X^d(t), \\
X^s(t)&=&\int_{-\infty}^\infty d\omega\,e^{-i\omega t}\mathbf{G}(\omega)\tilde\eta(\omega), \nonumber\\
X^d(t)&=&\int_{-\infty}^\infty d\omega\,e^{-i\omega
t}\mathbf{G}(\omega)\tilde F(\omega), \nonumber
\end{eqnarray} where
$\mathbf{G}(\omega)=(\mathbf{\Phi}-\omega^2\mathbf{M}-i\omega\mathbf{\Gamma})^{-1}$
is the phonon Green's function. It is clear that the displacement
vector $X(t)$ is a superposition of two contributions: $X^s$, as the
effect of the stochastic heat bath, and $X^d$, as the consequence of
the driving force. For our periodic driving case, it is easy to get
$X^d(t)=-\text{Im}[\mathbf{G}(\Omega)\tilde fe^{-i\Omega t}]$, with
$\tilde f_i=\delta_{i\alpha}A_0$. Similarly, the velocity has the
same decomposition $\dot{X}=\dot{X}^s+\dot{X}^d$, with
$\dot{X}^d=\text{Im}[i\Omega\mathbf{G}(\Omega)\tilde fe^{-i\Omega
t}]$. Straightforwardly, following the definition, the local energy
current can be similarly expanded as well:
\begin{eqnarray}\label{J_i}
I_i(t)&=&I^s_i(t)+I^d_i(t)\\
I^s_i(t)&=& -k\langle\dot x^s_{i}(x^s_{i}-x^s_{i-1})\rangle \nonumber\\
I^d_i(t)&=& -k\dot x^d_{i}(x^d_{i}-x^d_{i-1}). \nonumber
\end{eqnarray}
With $I_i(t)$ written in this form, it is obviously that the
contributions of noise term and driving force to the energy
current are independent from each other.
We should note this independence comes from the
fact that the driving force is not statistically correlated to the
heat baths. Since in the present case, the driving force is
deterministic, we have dropped the notation of ensemble average
$\langle\cdots\rangle$ in $I^d_i(t)$.

$I^s_{i}(t)$ is just the conventional steady-state heat flux, which
is proportional to the temperature difference and is given by
$I^s_{i}=\frac{\gamma^2(T_L-T_R)}{\pi}\int
d\omega\,\omega^2|G_{1N}(\omega)|^2$ ~\cite{Dhar2008}. Thus, the
contribution of stochastic heat bath to the oscillatory steady state
energy current is just
$J^s_i=\Omega/(2\pi)\int_0^{{2\pi}/{\Omega}}I^s_idt=I^s_i$. In all
the following discussions and simulations, we will set $\Delta T=T_L-T_R=0$
unless otherwise stated, since a finite temperature
difference mainly causes an additive shift of the energy current
curve.

To obtain the expression of $J^d_{i}$, we just need to calculate the
product of displacement and velocity contributed from the driving
force by applying Eq.~\eqref{X_D}. Denoting
$\mathcal{R}(\Omega)\equiv\text{Re}[\mathbf{G}(\Omega)]$ and
$\mathcal{I}(\Omega)\equiv\text{Im}[\mathbf{G}(\Omega)]$, we have
\begin{eqnarray}
x^d_{i}\dot
x^d_{j}=A_0^2\Omega[\mathcal{R}_{i\alpha}\mathcal{I}_{j\alpha}\sin^2(\Omega
    t)-\mathcal{I}_{i\alpha}\mathcal{R}_{j\alpha}\cos^2(\Omega
    t)\nonumber\\
    +\frac{1}{2}\mathcal{R}_{i\alpha}\mathcal{R}_{j\alpha}\sin(2\Omega t)-\frac{1}{2}\mathcal{I}_{i\alpha}\mathcal{I}_{j\alpha}\sin(2\Omega
    t)].
\end{eqnarray}
Since the driving force is periodic such that the steady state
energy current is oscillatory, we take the periodic average of the
physical quantities, and the final expression of
driving-force-contributed energy current is:
\begin{align}\label{j_Di_hm}
J^d_{i}&=\frac{kA_0^2\Omega}{2}[\mathcal{R}_{i-1\alpha}(\Omega)\mathcal{I}_{i\alpha}(\Omega)
-\mathcal{I}_{i-1\alpha}(\Omega)\mathcal{R}_{i\alpha}(\Omega)]\nonumber\\
&=\frac{kA_0^2\Omega}{2}\text{Im}[{G}^*_{i-1\alpha}(\Omega){G}_{i\alpha}(\Omega)].
\end{align}
Therefore, when $T_1=T_2$, $J_i=J^s_i+J^d_i=J^d_i$. In fact, $J^d_i$
has two values ($i\leqq\alpha$ and $i>\alpha$) and is constant on
each side of the driven site $\alpha$. A direct proof is
detailed in the Appendix.

From the expression of Eq.~(\ref{j_Di_hm}), we should be aware that
the energy flux possesses the denominator
$D(\Omega)\equiv|\det[\textbf{Z}(\Omega)]|^2$ with
$\textbf{Z}(\Omega)\equiv
\mathbf{\Phi}-\Omega^2\mathbf{M}-i\Omega\mathbf{\Gamma}$. Therefore,
when $D(\Omega)$ approaches its minimums under certain driving
frequencies, the energy flux will exhibit its maximum values. As a
consequence, resonance emerges. Let us denote
$P_N(\Omega)=\det(\mathbf{\Phi}_N-\Omega^2\mathbf{M})$ to be the
characteristic polynomial of the $N \times N$ force matrix
$\mathbf{\Phi}_N$ with $N$ particles. It can be shown that,
\begin{align}
    \det[\mathbf{Z}(\Omega)]=P_N(\Omega)-\gamma^2\Omega^2P_{N-2}(\Omega)-2i\gamma\Omega
    P_{N-1}(\Omega),
\end{align}
where $P_{N-1}(\Omega)$ is the characteristic polynomial of the
$(N-1) \times (N-1)$ force matrix $\mathbf{\Phi}_{N-1}$ with the
first row and column or the last row and column taken out from
$\mathbf{\Phi}_N$. $P_{N-2}(\Omega)$ is the characteristic
polynomial of the $(N-2) \times (N-2)$ force matrix
$\mathbf{\Phi}_{N-2}$ with both the first and last rows and columns
taken out from $\mathbf{\Phi}_{N}$. Therefore, the denominator
$D(\Omega)$ is given by
\begin{equation}\label{denomn}
    D(\Omega)=[P_N(\Omega)-\gamma^2\Omega^2
    P_{N-2}(\Omega)]^2+4\gamma^2\Omega^2P_{N-1}^2(\Omega).
\end{equation}
Apparently, resonant frequencies correspond to those values of
$\Omega$ which make $D(\Omega)$ locally minimized. It is difficult
to get the explicit exact solutions for those locally minimized
solutions. Nevertheless, much simpler and inspiring results can
still be obtained if we consider the limiting case with either small
friction coefficient or sufficiently large friction. For small
friction, we can just set $\gamma = 0$ in Eq.~\eqref{denomn} and
obtain $D(\Omega)= P_N^2(\Omega)$. Then, the resonant frequencies of
the energy current are just the $N$ eigenfrequencies of the force
matrix $\mathbf{\Phi}_N$. For large friction, by keeping only the
highest order terms of $\gamma$ in Eq.~\eqref{denomn}, we get
$D(\Omega)\sim P_{N-2}^2(\Omega)$. In this limiting case, the
resonant frequencies correspond to the $N-2$ eigenfrequencies of
$\mathbf{\Phi}_{N-2}$. When the friction is in between the two
cases, we have to minimize $D(\Omega)$ to get those resonant
$\Omega$, which should be a gradual shift between the
eigenfrequencies of $\mathbf{\Phi}_{N-2}$ and $\mathbf{\Phi}_{N}$.

\begin{figure}
\includegraphics[scale=0.34]{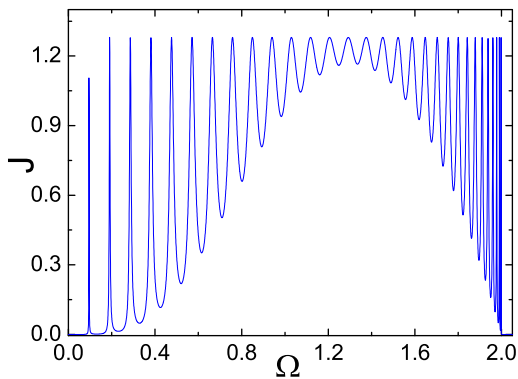}
\caption{\label{fig:J_hm} (Color online) Steady state energy current
versus driving frequency in the harmonic lattice. $k=1$, $k_o=0$,
$A_0=3.2$, $\gamma=1$, $\Delta T=T_L-T_R=0$, $N=32$ and $\alpha=1$.}
\end{figure}

Using Eq.~\eqref{j_Di_hm}, we calculate energy current vs driven
frequencies, as shown in Fig.~\ref{fig:J_hm}. It is not surprising
to observe the multiple-peak resonance behavior in the harmonic
case, since whenever the driven frequency approaches one of the
system's eigenfrequencies, resonance will occur. The number of the
eigenfrequencies is mainly determined by the system size.

We can also define other quantities such as the rate of heat
released from the two heat baths and rate of work done by the
driving force. Their definitions and analytical expressions are
given below:
\begin{equation}\label{qL_hm}
\dot q_L\equiv\langle\dot x_1(\eta_L-\gamma\dot x_1)\rangle=\langle
\dot x^s_{1}(\eta_L -\gamma\dot x^s_{1})\rangle-\gamma (\dot
x^d_{1})^2,
\end{equation}
where $\dot q_L$ stands for the rate of heat released from the left
bath. The first term is contributed by heat baths and is zero when
$\Delta T=0$, while the second term comes from the driving force.
Therefore the periodic average at $\Delta T=0$ gives
\begin{equation}\label{dot_q_L}
\dot Q_L= \frac{\Omega}{2\pi}\int_0^{{2\pi}/{\Omega}}dt \dot q_L = \dot Q^s_L + \dot Q^d_L = -\frac{\gamma
A_0^2\Omega^2}{2}|G_{1\alpha}(\Omega)|^2.
\end{equation}
Similarly, the expression for $\dot Q_R$, the average rate of heat
released from the right bath, is given as follows
\begin{equation}\label{q_R_dot}
\dot Q_R=  \dot Q^s_R + \dot Q^d_R = -\frac{\gamma
A_0^2\Omega^2}{2}|G_{N\alpha}|^2.
\end{equation}
The rate of work done by the driving force is defined as
\begin{equation}
\dot w(t)\equiv\langle f(t)\dot x_\alpha\rangle=f(t)\dot
x^d_{\alpha}
\end{equation}
where $f(t)=A_0\sin(\Omega t)$. Its periodical average gives
\begin{equation}\label{w}
\dot W=\frac{\Omega}{2\pi}\int_0^{{2\pi}/{\Omega}}dt \dot w(t)=\frac{A_0^2\Omega}{2}\mathcal{I}_{\alpha\alpha}(\Omega).
\end{equation}
Just like the energy flux, $\dot W,\ \dot Q_L$ and $\dot Q_R$ also
show multi-resonance behavior \cite{zhangsongNUS}. By the continuity
equation, the energy current flowing out and into a particle must
cancel each other when the system reaches its steady state, since
the local energy density does not vary with time at steady state.
Therefore, we immediately obtain the relation $\dot W+\dot
Q_L=J_i=-\dot Q_R$ ($i>\alpha$), which obeys the energy
conservation. For the harmonic model, a direct proof is given in the
Appendix. We should also be aware that this relation also holds in
the FK model and this is verified by numerical results.

\subsection{Multiple resonances of energy transport in force-driven FK model}

\begin{figure}
\includegraphics[scale=0.34]{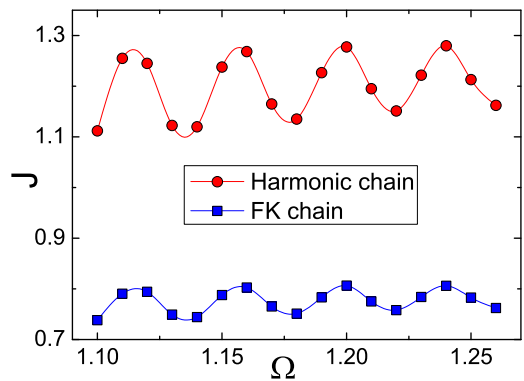}
\caption{\label{fig:zoom_in} (Color online) Comparison of the FK
case and harmonic case with force amplitude $A_0=3.2$ in a randomly
selected frequency region. Other parameters are $k=1$, $k_o=0$,
$V=5$, $\gamma=1$, $\Delta T=0$, $N=64$ and $\alpha=1$.}
\end{figure}

The main qualitative difference between the multiple-resonance
curves in Fig.~\ref{fig:MRP} and~\ref{fig:J_hm} lies in the fact
that the height of the peak for the FK case is smaller than that for
the harmonic case, due to the nonlinear on-site potential. This is
more obvious for low and high frequency regime. More importantly, we
notice that the positions of multiple peaks of the two models seem
to be close to each other, as illustrated in Fig. \ref{fig:zoom_in}.
This suggests that the FK model and harmonic model share a similar
resonance mechanism. The harmonic model discussed in
Sec.~\ref{sec:A} is able to shed lights on the multi-resonance
mechanism in a force-driven nonlinear lattice. It is worthwhile to
point out that when sampling frequency is not dense enough, one may
get a single peak, which actually corresponds to the lower envelope
of the multi-resonance curve. When $A_0$ decreases, the system seems
to have a single peak even though we increase the density of the
sampling points, as shown in Fig.~\ref{fig:MRP}. The mechanism for
the single peak should still be explained by the fact that the
driving frequency is resonant with the system's characteristic
frequencies. However, the non-linear potential smooths out the peaks
of multi-resonances. In this way, we can only observe the lower
envelope so that only one peak is observable. Thus the
single-resonance curve does not completely reflect the intrinsic
transport property of the FK model.

\begin{figure}
\includegraphics[scale=0.36]{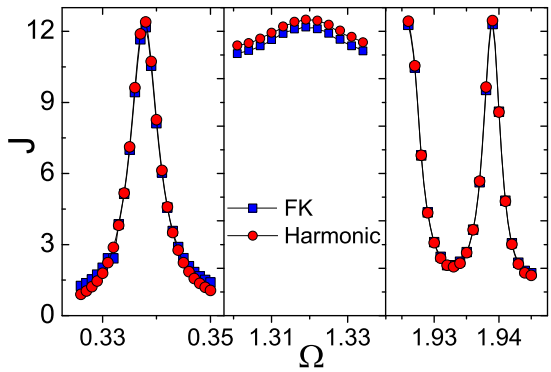}
\caption{\label{fig:F10} (Color online) Heat current vs driven
frequency for large force amplitude $A_0=10$. (a) Low frequency
regime. (b) Moderate frequency regime. (c) High frequency regime.
Other parameters are the same as in Fig.~\ref{fig:zoom_in}.}
\end{figure}

Let us examine how the parameters of the system will affect the
multi-resonance curve. We have already seen the case of the
appearance of multiple peaks when $A_0$ increases, as shown in
Fig.~\ref{fig:MRP}. When $A_0$ becomes large, the kinetic energy of
the particle gained from driving force will be much larger than the
height of the on-site potential. In this way, the mask effect of
nonlinear on-site potential becomes negligible and the FK model
effectively reduces to a harmonic model without on-site potential.
This is also demonstrated in Fig. \ref{fig:F10} with $A_0=10$. We
arbitrarily choose three typical regimes of driving frequency:
small, moderate and large $\Omega$. In all three regimes, the FK
model is very close to the harmonic one. Thus, as long as $A_0$
becomes large enough, the mask effect of nonlinearity will fade away
and the intrinsic multiple resonant peaks will emerge.

\begin{figure}
\includegraphics[scale=0.36]{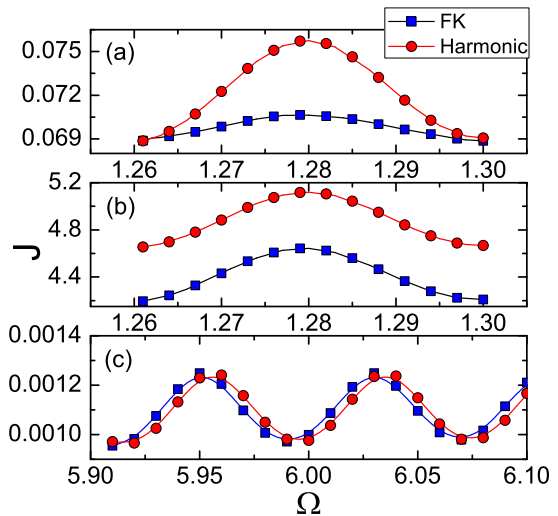}
\caption{\label{fig:T} (Color online) Temperature effect on
comparison of the FK case and harmonic case. (a) $T=2.0$, $A_0=1$.
(b) $T=0.01$, $A_0=6.4$. (c) $T=0.01$, $A_0=0.1$. Other parameters
are the same as in Fig.~\ref{fig:zoom_in}, except in (c): $k=10$,
$k_o=15$ and $V=15$.}
\end{figure}

We are also interested in the effect of temperature. Considering
that $T$ (defined as $T\equiv {(T_L+T_R)}/{2}$) will take the role
of thermal excitation which mainly affects the kinetic energy of
particles, we expect a large value of $T$ will have a similar effect
with $A_0$ on the resonance behaviors. As shown in
Fig.~\ref{fig:MRP} with $T=0.5$ and $A_0=1$, there is no multiple
resonances behavior. While for the increased temperature case
$T=2.0$, there is clearly a peak in a small $\Omega$ range, as shown
in Fig.~\ref{fig:T}(a), which indicates the multiple peaks resonance
behavior in the whole frequency range. At low temperature with large
amplitude, this is also the case, as illustrated in
Fig.~\ref{fig:T}(b). When $T$ and $A_0$ are both small, compared to
the nonlinear on-site potential $V$, particles are confined near
their equilibrium positions, within the valley of non-linear on-site
potential. By Taylor expansion of the potential to the second order,
we see that the non-linear potential reduces to a harmonic one, so
that the harmonic approximation works. In this case, the lower bound
of the phonon band will be raised by $\sqrt V$. The phonon band in
this case is therefore moved from $0 < \Omega < \sqrt{4k}$ to $\sqrt
V < \Omega < \sqrt{V+4k}$ \cite{interface}. (Accordingly, the phonon
band of the harmonic model is shifted by the on-site potential $k_o$
to $\sqrt k_o < \Omega < \sqrt{k_o+4k}$). Energy transport is
forbidden outside the frequency region. It is thus expected to see a
shift of the energy flux curve to the right while the
multi-resonance are still observed, as shown in Fig.~\ref{fig:T}(c).

The above discussion reveals that there are mainly three dynamic
regimes:

(1) When $k_BT+{A_0a}/({2\pi})\gg{V}/{(2\pi)^2}$ ($a$ is
the lattice constant as defined previously), the FK model will
approach a harmonic model \emph{without} an on-site potential. Multi-resonances are observable.

(2) For the opposite case, in which
$k_BT+{A_0a}/({2\pi})\ll{V}/{(2\pi)^2}$, the FK model reduces to a
harmonic model \emph{with} pure harmonic on-site potential of
strength $k_o=V$. Multi-resonances are still observable.

(3) When $T+{A_0a}/({2\pi})$ is comparable with ${V}/{(2\pi)^2}$,
the multiple peaks will be smoothed out by the effect of nonlinear
on-site potentials and only single resonant peak is observable. In
the intermediate regimes, there is the crossover from single peak to
multiple peaks, of which the resonant magnitudes are smaller
compared with those in the harmonic model.

In addition, the heat released from the baths and the work done by
the force in FK model also show parameter-dependent multiple
resonances \cite{zhangsongNUS}, which can be explained by the same
mechanism.

\section{Absence of heat pumping in a force-driven lattice}

\begin{figure}
\includegraphics[scale=0.33]{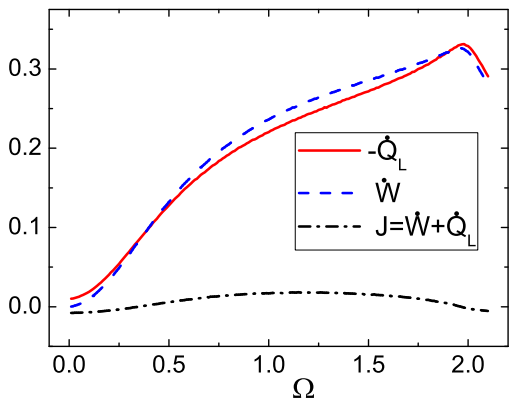}
\caption{\label{fig:fig7}  (Color online) $-\dot{Q}_L$, $\dot{W}$,
$J$ vs the driving frequency in force-driven FK model. Parameters
are the same as those for lower single peak curve with $A_0=1$,
$T_L<T_R$ in Fig.~\ref{fig:MRP}. $\dot{Q}_L$ is negative at all
frequencies, showing a negative result for pumping effect. }
\end{figure}

\begin{figure}
\scalebox{0.85}[0.75]{\includegraphics{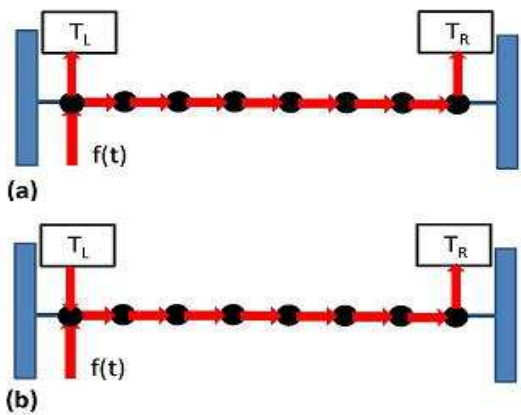}}\vspace{-4mm}
\caption{\label{fig:pump} (Color online) (a) A typical energy flow
diagram in force-driven FK model with $T_L < T_R$. (b) The situation
when a model performs as a heat pump with $T_L < T_R$. The arrow
denotes the direction of the energy flow.}
\end{figure}

In all simulations so far, we only consider the energy flux flowing
into the right reservoir as a whole. In fact it is a sum of two
parts: the rate of heat released from the left reservoir into the
system and the rate of work done by $f(t)$, \emph{i.e}, $J=-\dot Q_R=\dot Q_L
+ \dot W$. Now let's consider a typical result of energy transport
in a force-driven FK lattice with $T_L<T_R$, where these two
contributions are separated, as shown in Fig.~\ref{fig:fig7}.
Notably, though $J$ is positive (from $L$ to $R$) in the resonance
region, $\dot Q_L$ is negative in the full range. This indicates
heat will always flow into the left reservoir from the chain,
whatever the driving frequency is. The corresponding energy flow
diagram is depicted in Fig.~\ref{fig:pump}(a). This actually shows
that the FK system fails to function as a heat pump since normally,
a pump should have an energy flow diagram as in
Fig.~\ref{fig:pump}(b), where the heat is absorbed from the low
temperature bath and released to the high temperature one. We have
tested a broad range of parameters and even in the regimes with
multiple resonances. In all cases, the energy never flows as in
Fig.~\ref{fig:pump}(b). Therefore we conclude that there is no heat
pumping action in a force-driven FK lattice.

This finding is consistent with Ref.~\cite{Marathe07}, wherein it is
found that two coupled oscillators of either harmonic or FPU-like
interaction would not act as a heat pump under external driving
forces. Ref.~\cite{Ai2010} gets the opposite conclusion, because
they overlook the direction of $\dot Q_L$ and claim the model can
act as a heat pump whenever $J$ flows to the right bath of a higher
temperature.

To understand the underling mechanism for the absence of heat
pumping in a force-driven lattice system, we may consider
Eq.~\eqref{qL_hm}. The first stochastic contribution is always
negative since $T_L<T_R$. And the second driving contribution, which
does not depend on the
temperature, always contributes a negative value as well. 
Thus, $\dot Q_L$ will still be negative, which indicates the low
temperature bath always absorbs rather than releases heat, even we
may change the position of the driving force. The situation just
corresponds the energy flow diagram in Fig.~\ref{fig:pump}(a).

Clearly as long as the effects of the driving force and the baths
are independent, Eq.~\eqref{qL_hm} holds for harmonic chain. Thus
for a harmonic system, in order to get a heat pump, a necessary
condition is that the driving force is statistically correlated to
the bath, such that these two contributions could be synergetic.
While in the FK model, effects from the deterministic driving
force and stochastic baths are not separable as in Eq.~\eqref{qL_hm}
because of the nonlinear on-site potential. However, the nonlinearity can not
yet make the two contributions synergetic. And still, our numerical
simulations show a negative result for a force-driven FK lattice
acting as a heat pump. We thus speculate that for both the
force-driven harmonic and nonlinear lattice, the correlation may be
the key ingredient for the presence of heat pump action, similar to
the entanglement in quantum thermal baths \cite{single}, the
off-equilibrium nonthermal reservoirs \cite{Das} and the bath noise
correlation in temperature-driven case \cite{Ren10}, for ``low
temperature to high temperature'' thermal transports.

\section{Conclusion}

To summarize, we have studied the energy transport control in a
force-driven one dimensional nonlinear lattice--the
Frenkel-Kontorova model. We have found multiple thermal resonances
as a function of the driven frequency. Although not exactly the
same, the resonant frequencies are closely related to the
eigenfrequencies of the force matrix, and the number of resonant
peaks is bounded above by the system size $N$. Moreover, since the
onsite nonlinear potential tends to decrease the magnitude of energy
current and smoothes out the multiple peaks, the multi-resonance
phenomenon is only observable in certain parameter ranges and the
crossover from multiple resonances to single resonance will occur.
Finally, by following a rigorous definition of heat pump, we clarify
a previous contradiction and conclude that heat pumping effect is
absent in force-driven lattices.

In this paper, we focus on a one-dimensional chain. However the
analysis and results can be generalized to high dimensional system
with arbitrary topology, like polymer networks, proteins
\cite{network1}, and the three dimensional random elastic network
\cite{network2}, of which the eigen-spectra are much more
complicated and non-trivial. The investigation of the resonances in
such systems will give us more flexible methods for mechanical
control of energy transport in real applications.

\section{Acknowledgement}
This work has been supported in part by an NUS grant, R-144-000-285-646. This work results from Mr. Zhang Song's ``UROPS'' (Undergraduate
Research Opportunities) project under supervision of Jie Ren and Baowen Li.
\section*{APPENDIX}
Intuitively, from the energy conservation point of view, the
following relations should hold
\begin{align}
-\dot Q_R&=J_{\alpha+1}=\cdots=J_N,\label{eq:qRJ}\\
\dot Q_L&=J_1=\cdots=J_\alpha,\label{eq:qLJ}\\
\dot W&=-\dot Q_L-\dot Q_R. \label{eq:qw}
\end{align}
However, it is highly nontrivial to give direct proofs for arbitrary
nonlinear force-driven lattices. In the following, we will provide a
direct proof of the above equations in force-driven harmonic chains
by using the analytical forms of $\dot Q_{L,R}$, $\dot W$ and $J_i$.

Let us first define the dimensionless matrix
$\mathbf{\hat{Z}}(\Omega)\equiv {\mathbf{Z}(\Omega)}/{k}$, such that
${\hat{Z}}_{ij}=\delta_{i,j}(c-ib\delta_{i,1}-ib\delta_{i,N})-\delta_{i,j+1}-\delta_{i,j-1}$,
where $b={\gamma\Omega}/{k}$ and $c=2-{m\Omega^2}/{k}$ are two
dimensionless parameters. Here we set $k_o=0$ without loss of
generality. Considering $\mathbf{G}=\mathbf{Z}^{-1}={\mathbf{\hat
Z}^{-1}}/{k}$, we need a formula for the inversion of the matrix
$\hat{\mathbf{Z}}$, which is given in its recurrence
form~\cite{FON}:
\begin{align}
(\mathbf{\hat{Z}}^{-1})_{ij}=\frac{\theta_{i-1}\phi_{j+1}}{\theta_N}\quad\text{if
}
i\leq j,\label{eq:inv1}\\
(\mathbf{\hat{Z}}^{-1})_{ij}=\frac{\theta_{j-1}\phi_{i+1}}{\theta_N}\quad
\text{if } i>j,\label{eq:inv2}
\end{align}
where $\theta_i$ and $\phi_i$ are two sequences, defined by the
recurrence relations:
\begin{align}
\theta_i&={\hat{Z}}_{ii}\theta_{i-1}-\theta_{i-2},\label{eq:recc}\\
\phi_i&={\hat{Z}}_{ii}\phi_{i+1}-\phi_{i+2},\label{eq:recc2}
\end{align}
with the initial condition $\theta_0=1$, $\theta_1={\hat{Z}}_{11}$,
$\phi_{N+1}=1$ and $\phi_N={\hat{Z}}_{NN}$. Note $\theta_N$ is
actually the determinant of the matrix $\mathbf{\hat{Z}}$, which
will be denoted as $d$ in follows. Now, to prove Eq.~\eqref{eq:qRJ},
let us recall the expression for $\dot Q_R$ and $J_i$ (see
Eq.~\eqref{j_Di_hm} and~\eqref{q_R_dot}) and use Eq.~\eqref{eq:inv1}
and~\eqref{eq:inv2}. We then get (set $A_0=1$ for clearness):
\begin{align}
-\dot Q_R&=\frac{\gamma\Omega^2}{2}|G_{N\alpha}|^2
=\frac{\gamma\Omega^2}{2|d|^2k^2}|\theta_{\alpha-1}|^2,\\
J_i&=\frac{k\Omega}{2}\text{Im}[G_{i-1\alpha}^*G_{i\alpha}]\nonumber\\
&=\frac{\Omega}{2k|d|^2}\text{Im}[(\theta_{\alpha-1}\phi_i)^*(\theta_{\alpha-1}\phi_{i+1})]\nonumber\\
&=\frac{\Omega\text{Im}[\phi_i^*\phi_{i+1}]}{2k|d|^2}|\theta_{\alpha-1}|^2
\end{align}
where $i > \alpha$. Hence, to prove the above two equations are
equivalent with each other, we need to show
$\text{Im}[\phi_i^*\phi_{i+1}]=\gamma\Omega/k$ for all particles to
the right of the driving force. We do induction on the index $l$. For
$l=N$, the relation is clearly true, which can be taken as the base
case. Then suppose the relation is true for all $l \geq i$, we need
to show it is also true for $l=i-1$ case. Use the recurrence
equation Eq.~\eqref{eq:recc}, we have
$\text{Im}[\phi_{i-1}^*\phi_i]=\text{Im}[(c\phi_i^*-\phi_{i+1}^*)\phi_i]
=-\text{Im}[\phi_{i+1}^*\phi_i]=\text{Im}[\phi_i^*\phi_{i+1}]=\gamma\Omega/k$.
Thus, we have finished the proof that Eq.~\eqref{eq:qRJ} holds for
all particles to the right of the force. A similar proof can be
given to show Eq.~\eqref{eq:qLJ} holds for all particles to the left
of the driving force.

Before proceeding to prove Eq.~\eqref{eq:qw}, we need more
notations. Denote the $n\times n$ matrix $\mathbf{M^{(n)}}$ with
elements
${M^{(n)}}_{ij}=c\delta_{i,j}-\delta_{i,j+1}-\delta_{i,j-1}$ and its
determinant $K_n=det[\mathbf M^{(n)}]$, with boundary $K_{-1}=0$ and
$K_0=1$. The explicit expression of the determinant is given by
$K_{n-1}={(y_1^n-y_2^n)}/{(y_1-y_2)}$, where $y_1$, $y_2$ is the
solution of the quadratic equation $y^2-cy+1=0$. We observe that
$\theta_i$ is in fact the determinant of the sub-matrix in
$\mathbf{\hat{Z}}$ starting from the $1st$ row and column to the
$ith$ row and column. Similarly $\phi_i$ is the determinant of the
sub-matrix in $\mathbf{\hat{Z}}$ starting from the $ith$ row and
column to the $Nth$ row and column. Now we may expand $\theta_i$,
$\phi_i$ and the determinant of $\mathbf{\hat{Z}}$ in terms of $K_n$
\begin{align}
d&=K_N-b^2K_{N-2}-2ibK_{N-1},\\
\theta_i&=K_i-ibK_{i-1},\\
\phi_i&=K_{N-i+1}-ibK_{N-i}.
\end{align}
Now we are ready to prove Eq.~\eqref{eq:qw}. Let us assume
Eq.~\eqref{eq:qw} holds and substitute in the expressions of $\dot
W$, $\dot Q_L$ and $\dot Q_R$ (see
Eq.~\eqref{dot_q_L},~\eqref{q_R_dot} and~\eqref{w}). Then
Eq.~\eqref{eq:qw} becomes
$\mathcal{I}_{\alpha\alpha}=\gamma\Omega(|G_{1\alpha}|^2+|G_{N\alpha}|^2)$.
With $b={\gamma\Omega}/{k}$ and $G={\mathbf{\hat{Z}}^{-1}}/{k}$,
also using Eq.~\eqref{eq:inv1}$\sim$~\eqref{eq:recc2}, we get
\begin{equation}\label{eq:ted}
\text{Im}(\theta_{\alpha-1}\phi_{\alpha+1}d^*)=b(|\theta_{\alpha-1}|^2+|\phi_{\alpha+1}|^2).
\end{equation}
Express the left hand side (LHS) and right hand side (RHS) of
Eq.~\eqref{eq:ted} in terms of $K_n$ (see Eq.($25$)$\sim$($27$)),
and after some algebraic work, it can be shown
\begin{align}
\text{LHS}=&b^3(K_{\alpha-2}K_{N-2}K_{N-\alpha}+K_{\alpha-1}K_{N-2}K_{N-\alpha-1}\nonumber\\
&-2K_{\alpha-2}K_{N-1}K_{N-\alpha-1})\nonumber\\
&+b(2K_{\alpha-1}K_{N-1}K_{N-\alpha}-K_{\alpha-2}K_{N}K_{N-\alpha}\nonumber\\
&-K_{\alpha-1}K_{N}K_{N-\alpha-1}),\\
\text{RHS}=&b^3(K_{\alpha-2}^2+K_{N-\alpha-1}^2)+b(K_{\alpha-1}^2+K_{N-\alpha}^2),
\end{align}
Now in order to show Eq.~\eqref{eq:qw} holds, it is sufficient to
prove LHS=RHS to complete the proof. Denote the coefficient of $b^3$ ($b$) in
LHS by $L_3$ ($L_1$) and that of RHS by $R_3$ ($R_1$). By using the
formula $K_{n-1}={(y_1^{n}-y_2^{n})}/{(y_1-y_2)}$ and the fact
$y_1y_2=1$, we get
\begin{align}
L_3=&\frac{1}{(y_1-y_2)^3}[y_1^{2\alpha-1}-y_2^{2\alpha-1}+y_1^{2N-2\alpha+1}-y_2^{2N-2\alpha+1}\nonumber\\
&-(y_1^{2\alpha-3}-y_2^{2\alpha-3})
-(y_1^{2N-2\alpha-1}-y_2^{2N-2\alpha-1})\nonumber\\
&-4(y_1-y_2)]=R_3.
\end{align}
Note that by a relabeling of $\alpha\rightarrow\alpha+1$ and
$N\rightarrow N+2$ in $L_3$ and $R_3$, and utilizing the relation
$K_n=cK_{n-1}-K_{n-2}$, we can show $L_3\rightarrow L_1$ and
$R_3\rightarrow R_1$. Therefore, LHS=RHS and the proof is completed.

\end{document}